\documentclass[aps,prb,floats,onecolumn,showpacs,preprintnumbers,10pt]{revtex4}
\usepackage{natbib}
\usepackage{amsmath}
\usepackage{amsfonts}
\usepackage{amssymb}
\usepackage{graphicx}
\usepackage{dcolumn}
\usepackage{bm}
\usepackage{subfigure}
\usepackage{textcomp}
\usepackage{enumerate}

\begin{document}
\title{\textit{Quantum effects in proton-conducting oxides: an exhaustive study in barium stannate.}}
\author{Gr\'egory Geneste$^1$}
\email{gregory.geneste@cea.fr}
\author{Alistar Ottochian$^{2}$} 
\author{Jessica Hermet$^{1,2}$}
\author{Guilhem Dezanneau$^{2}$}
\affiliation{$^1$ CEA, DAM, DIF, F-91297 Arpajon, France  \\
$^2$ Laboratoire Structures, Propri\'et\'es et Mod\'elisation des Solides, UMR CNRS 8580, \'Ecole Centrale Paris, Grande Voie des Vignes, 92295 Ch\^atenay-Malabry Cedex, France }
\date{\today}

\begin{abstract}
Density-functional theory calculations are performed to investigate hydrogen transport in the proton conductor BaSnO$_3$. Structural optimizations in the stable and saddle point configurations for transfer and reorientation allow description of the high-temperature classical and semi-classical regimes, in which diffusion occurs by over-barrier motion. At lower temperature (typically below 300 K), we describe a thermally-assisted quantum regime. In this regime, transfer and reorientation occur when the surrounding matrix adopts particular "coincidence" configurations in which quantum tunneling is favored. Both the non-adiabatic and the adiabatic cases are examined. In the adiabatic case, the energy landscape of hydrogen in the coincidence configuration is very flat, with very low coincidence energy barriers. Path-integral molecular dynamics simulations of the H atom in the coincidence potential reveal, in the transfer case, highly quantum behavior up to T=300~K (the density of probability of H in the coincidence configuration, has its maximum at the saddle point, due to the fact that the zero-point energy exceeds the coincidence energy barrier). Arguments are given that support the adiabatic picture for the transfer mechanism. This suggests existence of this state of hydrogen during the very short lifetime of the coincidence configurations ($\sim$ 10$^{-13}$ s), as a transition state for the transfer mechanism. Remarkably, such state is identical to that of ice X, a highly quantum phase of ice observed at high pressures $\approx$ 100 GPa. In the case of reorientation, typical times for existence of the coincidence configuration and for protonic motion are roughly equal, suggesting that the adiabatic picture is not valid. Protonic transfer and reorientation in oxides are therefore governed by radically different mechanisms below room temperature.
\end{abstract}

\maketitle


\section{Introduction}
The diffusion of light particles in condensed matter has always been a challenge for theory, especially in the regimes where the light particle behaves quantum mechanically, in a surrounding medium constituted by heavier atoms that rather behave classically. The pioneering ideas have been formulated by Marcus to describe the transfer of localized electrons in solutions~\cite{marcus1,marcus2}. A fundamental idea is that in such regime, transfer is rather of quantum nature (tunneling) and is controlled, not by the thermal fluctuations of the light particle itself (which is frozen in its ground state), but by those of the surrounding atoms. Thermal fluctuations may occasionally create atomic configurations in which the light particle has the same energy when placed on both sides of the barrier, making in coincidence its energy levels and thus enhancing tunneling probability, with an energy barrier generally lower and less wide than in the stable configuration. The configurations in question can be called "coincidence configurations". These concepts, that hold for localized electrons, can be, to a certain extent, generalized to light atoms such as hydrogen~\cite{stoneham1,stoneham2}.

Hydrogen diffusion is a key process in a large number of systems, organic or inorganic. As an impurity occupying interstitial positions in a solid, H can exist under several chemical forms, due to its medium electronegativity (proton H$^+$, neutral atomic H or molecular H$_2$, or hydride H$^-$/H$^{\delta -}$). Protonic transfer, in particular, is the central mechanism of acid-base reactions. Technological applications include solid electrolytes for polymer exchange membrane fuel cells (PEMFC) or protonic ceramic fuel cells (PCFC), sensors, gas separation membranes, electrolyzers or metal hydrides for hydrogen storage.

Hydrogen is the lightest element of the periodic table, and as such, exhibits behavior that significantly differs from that of the other, heavier, atoms. First, H has a faster dynamics, and in some cases, can be considered as adjusting instantaneously its state to that of its environment. Second, owing to this light mass, the quantum nature of hydrogen motion is manifest up to temperatures higher than for the other atoms (the magnitude of its zero-point energy is higher, and the eigenstate quantization of its vibration modes is stronger than for the other, heavier, atoms).

In a host crystal, a hydrogen impurity should thus, {\it a priori}, be considered as behaving according to the laws of quantum mechanics rather than those of classical mechanics. Moreover, the quantum nature of hydrogen should survive up to temperatures higher than for the other atoms.  As an example, the energy quantum associated with the stretching vibration of the OH group in BaZrO$_3$ (BZO) is $\hbar \omega$ = 435 meV~\cite{sundell2007}, corresponding to a characteristic temperature $\hbar \omega / k_B \approx$ 5000 K. The two modes related to lateral vibrations of OH in this compound~\cite{sundell2007} also have high energies of 110 and 75 meV (1275 and 870 K). Thus at room temperature, the proton in BZO, and in many other oxides, should be, {\it as long as harmonic behavior is maintained}, completely frozen in its vibrational ground state.

In the periodic potential of a crystal, a quantum particle is supposed to behave as a Bloch wave, and thus exhibit a density of probability having the periodicity of the crystal. In its ground state, in particular, it should be totally delocalized over the crystal. However, in the case of the hydrogen atom, such delocalization does not occur, mainly for three reasons:

(i) as temperature increases, thermal vibrations of the H atom coupled to the host, a quasi-classical, macroscopic, system, are responsible for a quantum/classical transition;

(ii) the tunneling matrix elements of H between its different stable sites is rather weak, making impossible the formation of a band of delocalized states.

(iii) the crystal potential is not rigid, and even at low temperature, strong deformation of the surrounding matrix takes place, resulting in hydrogen localization at one site. Such deformation strongly stabilizes the H atom at one particular site and, at the same time, breaks the periodicity of the matrix, making the occupied site energetically much more stable than the others. This phenomenon is called {\it "self-trapping"}.

\section{Hydrogen diffusion in solids}

\subsection {Temperature-evolution of the transition rate}

Depending on temperature, hydrogen transport in solids might exhibit different regimes~\cite{fukai}:

\begin{enumerate}

\item High-temperature classical regime

\item Intermediate-temperature "semi-classical" regime

\item Phonon-assisted tunneling regime

\item Low-temperature tunneling regime

\end{enumerate}

\subsubsection {Classical regime: } 

At very high temperature, for $T >> T_D$ ($T_D$ is the Debye temperature of the system), the classical approximation is satisfied. All the atoms including H behave as classical particles. Classical Transition State Theory (TST) describes well hydrogen diffusion in this regime. The hopping rate can be expressed as $k = k_0 e^{- E_a / k_B T}$. In the harmonic approximation, the activation energy $E_a$ and prefactor $k_0$ write:

\begin{equation}
E_a =  V_{saddle}  - V_{st},
\end{equation}

\begin{equation}
k_0 = \frac{\Pi_{i=1}^{3N} \nu_i}{\Pi_{i=1}^{3N-1} \nu_i^s},
\end{equation}

in which $V_{st}$ (resp. $V_{saddle}$) is the energy of the stable self-trapped (resp. saddle point) configuration, and $\{ \nu_i  \}$ (resp. $\{ \nu_i^s \}$) is the set of $3N$ (resp. $3N-1$) eigenfrequencies of stable vibration modes in the stable self-trapped (resp. saddle point) configuration. The quantity $E_a = V_{saddle} - V_{st}$ is called hereafter classical barrier. 
Due to the strong quantization of the vibrational levels of H, $T_D$ is generally very high (a few thousands K).

\subsubsection {Semi-classical regime: }

Below $T_D$ and above some crossover temperature $T_c$ (typically around room temperature or below), 
hydrogen diffusion is correctly modeled by Classical TST, provided quantum corrections are included in the activation energy and in the prefactor of the transition rate. 
In such regime, diffusion occurs via over-barrier motion (tunneling is negligible), but quantum effects of the H motion should be accounted for. Diffusion is well described by a reaction coordinate $\lambda$, chosen as {\it an internal parameter related to the coordinates of H}. This regime holds as long as H is not -- in the self-trapped geometry -- frozen in its vibrational ground state.

The transition rate follows Arrhenius behavior $k = k_0^Q e^{- E_a^Q / k_B T}$, with an activation energy $E_a^Q$ and a prefactor $k_0^Q$ (the superscript $Q$ stands for "quantum correction") given, in the harmonic approximation~\cite{sundell2007}, as

\begin{equation}
\nonumber
E_a^Q = \{ V_{saddle} + \sum_{i=1}^{3N-1} \frac{1}{2} h \nu_i^s \} - \{  V_{st}+ \sum_{i=1}^{3N} \frac{1}{2} h \nu_i \} 
\end{equation}

\begin{equation}
= E_a + \{ \sum_{i=1}^{3N-1} \frac{1}{2} h \nu_i^s - \sum_{i=1}^{3N} \frac{1}{2} h \nu_i \}
\end{equation}

\begin{equation}
k_0^Q = \frac{k_B T}{h} \frac{ \Pi_{i=1}^{3N} [1 - e^{-{ \frac{h \nu_i}{k_B T}  }}]}
{ \Pi_{i=1}^{3N-1} [1 - e^{-{ \frac{h \nu_i^s}{k_B T}  }}]},
\end{equation}

Note that for $k_B T >> h \nu_i~\forall i$, $k_0^Q$ tends to the classical prefactor $k_0$, and that for $k_B T << h \nu_i~\forall i$, $k_0^Q$ tends to $\frac{k_B T}{h}$. The transition rate, in this limit, expresses as $k = \frac{k_B T}{h} e^{- E_a^Q / k_B T}$ (Eyring formula).

In this semi-classical regime, the activation energy $E_a^Q$ is the classical barrier corrected from zero-point energies in the saddle point and classical configurations.

\subsubsection {Phonon-assisted tunneling regime: }

Below the crossover temperature $T_c$, $k_B T$ is not large enough to allow thermally activated H over-barrier motion: H does not undergo thermal fluctuations any more, but its quantum fluctuations are large, and tunneling becomes progressively, as T decreases, the dominant transfer mechanism.

However, in the self-trapped configuration, tunneling is not possible because (i) the energy barrier H has to overcome is large and wide and, (ii) if one considers that H tunneling occurs at fixed positions of the surrounding atoms, the final site has a higher energy than the initial one, owing to the atomic distortions associated to the self-trapping mechanism.

Tunneling of H becomes nevertheless possible through thermal vibrations of the environment.
These thermal fluctuations are able to bring occasionally the system in a set of specific configurations in which self-trapping is weakened. In these "coincidence configurations", H has the same energy whether it is placed, and relaxed, on one side of the barrier or on the other (under fixed positions of the surrounding atoms).

There is an infinite number of such coincidence configurations, namely, all the ones that exhibit certain symmetries. 
Let us call $V_{c}$ their energies. The quantity $E_c = V_{c} - V_{st}$ (possibly corrected by the zero-point energies in the two configurations, coincidence and self-trapped) is called the {\it coincidence energy} (hereafter noted $E_c$, or $E_c^Q$ if quantum zero-point corrections are accounted for). It is the energy that must be paid by thermal fluctuations to take the system in a coincidence configuration with energy $V_c$, and plays therefore the role of an activation energy. Among all these coincidence configurations, the most probable is the lowest-energy one. In a frozen coincidence configuration, we call $E_a^c$ (coincidence barrier) the energy barrier the H atom has to overcome to jump onto the neighboring site.

In this regime, the reaction coordinate $\lambda$ relevant to describe H transfer should be an {\it internal parameter related to the coordinates of the surrounding atoms}. The behavior of H in the coincidence configuration depends on the energy landscape felt by this atom in this configuration, namely the width and the height ($E_a^c$) of the coincidence barrier. According to this energy surface, a wide range of behaviors can be predicted, depending on the typical time the H atom takes for tunneling through the barrier compared to the lifetime of the coincidence configuration. Two extreme cases can be described:

(i) if the probability for tunneling is small during the typical time for coincidence, diffusion occurs through a non-adiabatic mechanism~\cite{fukai,sundell2007}. This is typically the case if the coincidence energy $E_c$ (uncorrected from zero-point contributions) is significantly smaller than $E_a$. Thus the energy barrier remains high in the coincidence configuration, and  the hydrogen atom has not the time to tunnel through the barrier during the time scale for coincidence. In that case, the phonon-assisted tunneling diffusion of H is described by the Flynn-Stoneham model~\cite{flynn}, in which the transition rate writes

\begin{equation}
k^{Non-Ad} = \frac{\sqrt{\pi} |J_0|^2}{2 \hbar \sqrt{E_c k_B T}} e^{- E_c /k_B T},
\end{equation}

$E_c$ being the coincidence energy (or $E_c^Q$ if corrected from zero-point contributions) and $J_0$ the bare tunneling matrix element for the proton~\cite{fukai}.

(ii) if the tunneling probability is large during the typical time scale for coincidence, diffusion occurs through an adiabatic mechanism, i.e. H has the time to adjust its state to that of its environment and tunnel through the barrier. Such probability depends on the characteristics of the energy barrier in the coincidence configuration, i.e. its width, and also its height compared to the zero-point energy of H. A low and thin barrier is likely to be favorable to an adiabatic behavior. In such case, the coincidence energy is expected to be close to the semi-classical barrier $E_c \approx E_a^Q$, and the transition rate takes the simple form

\begin{equation}
k^{Ad} = k_{D} e^{- E_c /k_B T},
\end{equation}

$k_D$ being an attempt frequency related to the Debye frequency of the host (not related to H), and $E_c$ being the coincidence energy (or $E_c^Q$ if corrected from zero-point contributions). Typically, $k_D^{-1} \sim $ a few 10$^{-13}$ s.
The configuration that is usually taken as the most favorable to this adiabatic behavior is the one corresponding to the fully relaxed classical saddle point geometry~\cite{fukai,sundell2007}, in which H is left free to evolve owing to its rapid motions.

Note that if the ground state energy in the coincidence configuration exceeds the coincidence barrier, motion of the proton in the coincidence potential might be very fast, ensuring an adiabatic behavior. Such motion would rather correspond to quantum over-barrier rather than tunneling (see next section).

The question is whether a non-adiabatic or an adiabatic tunneling mechanism prevails. This might depend on the system and on temperature~\cite{fukai}. If $T$ is sufficient high, so that many phonons of the host matrix are excited (the matrix is considered to behave classically), a wide range of coincidence configurations might be visited at thermodynamic equilibrium. Among these the ones favorable to non-adiabatic tunneling processes (at the lowest energies), those favorable to adiabatic processes (at higher energies), and all the ones inbetween can be found.

The probabilities of occurrence of the latter should be smaller according to Boltzmann distribution, but the tunneling probability of H in such configurations largely higher. Thus there is probably a competition between non-adiabatic processes (coincidence configurations visited more often but small tunneling probability) and adiabatic ones (coincidence configurations more rarely visited but large tunneling probability), depending on the value of the bare tunneling matrix element $J_0$.

The transition from the semi-classical regime to the phonon-assisted quantum-tunneling regime has been extensively studied in group-V bcc transition metals (V, Nb, Ta), both experimentally~\cite{qi1983,messer1986} and theoretically (using phenomenological potentials and density-functional calculations~\cite{sundell2004}). From a macroscopic point of view, in such systems, the over-barrier and thermally-assisted tunneling regimes are characterized by different activation energies, yielding a change of the slope in the Arrhenius plot of the H diffusion coefficient around $\approx$ 250 K. The activation energy at high temperature is the semi-classical one ($E_a^Q$), while the one at low temperature is the phonon-assisted tunneling one ($E_c < E_a^Q$). Similar behavior has been evidenced at Ni(001) surfaces~\cite{mattsson1995}. Quantum tunneling of hydrogen is also observed in semiconductors, such as silicon, below $\sim$ 70-80 K~\cite{herrero1995,herrero1997}, or boron-doped silicon~\cite{noya1997}. Tunneling is, however, expected to play a significant role up to room temperature~\cite{forsythe1998}.

\subsubsection {Low-temperature regime: }
At very low temperature, when thermal fluctuations of the matrix itself become so small that it behaves quantum mechanically (only a few phonons excited), hydrogen might diffuse by coherent tunneling or incoherent hopping~\cite{fukai}.

\subsection{Extrapolation of the previous concepts to proton-conducting oxides}

The impact of quantum fluctuations on hydrogen transport in proton-conducting oxides is a long-standing debate~\cite{gross1999,sundell2007,zhang2008}. At the working temperatures of Protonic Ceramic Fuel cells (600-900 K), the migration of protons occurs by over-barrier motion and tunneling is negligible, according to the Path-Integral simulations by Zhang et al.~\cite{zhang2008}. However, quantum effects remain large (delocalization of the wave packet $\Delta r \sim$ 0.10-0.15~\AA~\cite{hermet-thesis}), and zero-point energy corrections significantly modify the classical barriers: in BZO, density-functional calculations~\cite{sundell2007} showed that zero-point effects decrease the classical energy barriers (0.21 eV for transfer and 0.18 eV for reorientation) by 0.12 eV and 0.04 eV respectively. {\it PCFC electrolytes are therefore working in the semi-classical regime}.

However, hydrogen tunneling in proton-conducting oxides at lower temperatures has been suggested by a few experimental works, especially in cerates: Kuskovsky et al. have given evidence for tunneling processes below 85 K in BaCe$_{1-x}$Nd$_x$O$_{3 - \delta}$~\cite{kuskovsky1999}, and hydrogen tunneling was also suggested in BaCe$_{1-x}$Y$_x$O$_{3 - \delta}$~\cite{cordero2008}. Moreover, the Path-Integral simulations of Zhang et al.~\cite{zhang2008} on BZO suggest existence of such effects below $\sim$ 200 K. The crossover temperature $T_c$ between the semi-classical and thermally-assisted quantum regimes can be roughly estimated, in the case of a parabolic energy barrier~\cite{gillan,voth1989}, as $T_c = \frac{\hbar \Omega}{2 \pi k_B}$, where $\Omega$ is the (imaginary) pulsation of the unstable mode at the saddle point. Within the values given by Sundell et al.~\cite{sundell2007} in BZO, we have $T_c \approx$ 230 K, in agreement with the calculations of Zhang et al.~\cite{zhang2008}.

The work of Sundell et al.~\cite{sundell2007} provides important results that help to understand how the concepts prevailing in metals and semiconductors can be extended, or not, to oxides. These authors computed the energy required to create the coincidence configuration for transfer, both in the adiabatic and non-adiabatic cases, and found similar values of 0.19 eV (much larger than in metals~\cite{sundell2004,sundell2004c,mattsson1993,mattsson1995}), thus very close to the classical barrier $E_a$=0.21 eV, as a consequence of strong coupling between H$^+$ and its surrounding, polarizable, matrix. In the adiabatic case, they find an energy barrier in the coincidence configuration ($E_a^c$) as low as 0.02-0.03 eV, whereas in the non-adiabatic case, they find it much larger (0.41 eV). This leads to a very small tunneling probability in the non-adiabatic case and thus, a transition rate much smaller than in the adiabatic picture. The authors concluded that the adiabatic picture might be more relevant to describe protonic transfer in BZO in the phonon-assisted quantum regime.

\subsection{Adiabatic picture of protonic diffusion}
In the thermally-assisted quantum regime, the proton behaves adiabatically if it remains, at every time, in the ground state corresponding to the potential created by the current configuration of the surrounding atoms (considered as classical point-like particles), which means that it has the time to adjust instantaneously its state to the fluctuations of the surrounding matrix. Most of the time, H$^{+}$ is self-trapped and (quantum-mechanically) fluctuates around its stable position, but occasionally, thermal fluctuations bring the system in a configuration close to the coincidence, during a typical lifetime $\tau_{at} \sim$ 10$^{-13}$ s. The characteristic time for H to adjust its state is provided by quantum mechanics, i.e. by the time-dependent Schr\"odinger equation.

When a coincidence environment for the proton occurs, as the result of thermal fluctuations, H$^{+}$ can be considered as instantaneously evolving in a symmetric potential starting from an initial state spatially localized on one side of the barrier (the self-trapped state). If H$^{+}$ is initially trapped on one side of the barrier, it means that it is not in an eigenstate of the Hamiltonian any more. It is in a linear combination of eigenstates, and possesses therefore an energy extension $\Delta E$. Let us denote by $\Delta t$ the typical time scale for crossing through the coincidence barrier. An adiabatic behavior is obtained when $\Delta t$ is significantly smaller than $\tau_{at}$, the typical lifetime of the coincidence configuration. The typical time scale separating two consecutive tunneling events is the one needed for the proton to undergo significant time evolution of its wave packet, or to spread over the double-well potential, starting from a state spatially localized on one side of the barrier. Such time scale is provided by the {\it time-energy uncertainty relation}, $\Delta t . \Delta E \sim \hbar$, where $\Delta E$ is the energy extension of the initial localized state in the symmetric potential~\cite{cohen}.

If the energy $E$ of the proton is lower than the coincidence barrier $E_a^c$, diffusion through the barrier is possible by tunneling. The typical time scale for tunneling is then $\Delta t  \sim \hbar / \Delta E$. However, if the coincidence barrier $E_a^c$ is very low, the energy $E$ of the proton might exceed it (this is always the case if the ground state energy exceeds $E_a^c$). The proton could then diffuse over the barrier without tunneling, but here again, the typical time needed for the protonic wave packet to spread over the potential well is $\Delta t  \sim \hbar / \Delta E$. In both cases, an adiabatic behavior is obtained provided $\Delta t << \tau_{at}$, so that the proton has the time to adjust its state to the variations of its atomic environment.

\section{System studied and methodology}
In the present work, we focus on H transport in barium stannate BaSnO$_3$ (BSO), a perovskite with cubic structure (space group $Pm\bar{3}m$), that exhibits interesting levels of protonic conduction when hydrated after aliovalent (acceptor) doping on the Sn site~\cite{schober,murugaraj,wang2011,wang2012,wang2013,li2013}. This is a possible candidate as an electrolyte for PCFCs. Theoretical calculations using both phenomenological potentials~\cite{wang2009} and density-functional calculations~\cite{bevillon2014} have provided information about hydration mechanisms and behavior of point defects (oxygen vacancies, protons, trivalent dopants) in this oxide. Protons in such materials are localized at interstitial sites, and stabilized under the form of OH groups. They diffuse by the succession of intra-octahedral hopping (or transfer) and reorientation of the OH bond, a combination usually called the "Grotthuss" mechanism.

We first determine, in Sec.~\ref{results} the stable and saddle point configurations for transfer and reorientation in barium stannate, assuming all the atoms are classical particles. From this, we estimate the classical energy barrier $E_a$, for transfer and reorientation, relevant in the classical regime. Then, using the harmonic approximation, we compute quantum corrections relevant to the semi-classical regime. This gives access to the semi-classical energy barrier $E_a^Q$. Then, we examine two kinds of coincidence configurations, that should correspond to adiabatic and non-adiabatic behaviors in the phonon-assisted quantum regime. We obtain in both cases, and for the two mechanisms, an estimation of the coincidence energy $E_c$, and the coincidence barrier $E_a^c$.

In Sec.~\ref{pi}, we focus our attention on the adiabatic case. Under such assumption, the thermalized motion of H is simulated in the coincidence configuration, by performing Path-Integral Molecular Dynamics (PIMD) trajectories of H, with all the other atoms fixed at the positions of the (adiabatic) coincidence configuration. The state of H obtained, that fully includes quantum fluctuations, can be considered as close to the transition state for hopping/reorientation in the thermally-assisted quantum regime in an adiabatic picture. PIMD provides the density of probability of the proton in this coincidence potential $V^c_{ad}(\vec r)$ ($\vec r$ is the proton position), including thermal and quantum effects. We give evidence for a highly quantum behavior of H in the coincidence potential for transfer, related to the very low value of $E_a^c$. In the case of reorientation, the coincidence barrier is higher and yields different behavior for H.

Finally, we interpret the results on the basis of a simple one-dimensional, quartic double-well potential. This model also allows to provide rough estimation for the typical times for tunneling/quantum over-barrier motion in $V^c_{ad}(\vec r)$. By comparing these times to the typical time for coincidence, we suggest that the adiabatic picture is reasonably well justified in the case of transfer, probably not in the case of reorientation.


\section{Computational details\label{details}}
We have performed density-functional theory calculations~\cite{kohn-sham}, with the ABINIT code~\cite{abinit}, in the Projector Augmented-Wave (PAW) framework~\cite{Torrent2008}, using the Generalized Gradient Approximation (GGA-PBE)~\cite{pbe}. 10 electrons are treated in the valence for Ba ($5s^2$, $5p^6$, $6s^2$), 6 for oxygen ($2s^2$, $2p^4$), while for Sn, two atomic data have been used: the first one has 14 electrons in the valence ($5s^2$, $5p^2$ and the $4d^{10}$ semicore shell), while the second one only has 4 electrons ($5s^2$, $5p^2$, the $4d$ electrons being frozen in the core). We perform two kinds of calculations: structural optimizations (under various constraints, see hereafter), and Path-Integral Molecular Dynamics (PIMD) computations to simulate the quantum effects in the dynamics of the H atom. The 14-electron atomic data has been used in a few structural optimizations, while the 4-electron atomic data has been used for structural optimizations, and in PIMD simulations. The plane-wave cut-off is 25 Hartrees in the first case (when using the 14-electron atomic data), 18 Hartrees in the second case. The 14-electron atomic data has been used to provide validation of the 4-electron one by comparison of computed energy barriers.

We use a 2$\times$2$\times$2 BSO supercell, in terms of the 5-atom unit cell (40 atoms), in which a hydrogen interstitial is added. The supercell is maintained in a +1 charge state using a uniform compensating background, ensuring that H is in a protonic state H$^{+}$ (there is no additional dopant in the supercell to compensate the charge of H$^+$). Structural optimizations are performed until all the cartesian components of atomic forces are below 2.0$\times$10$^{-4}$ Ha/bohr ($\approx$ 0.01 eV/{\AA}). The first Brillouin zone of the supercell is sampled by a 3$\times$3$\times$3 {\bf k}-point mesh for the structural optimizations, and by a 2$\times$2$\times$2 {\bf k}-point mesh in the PIMD simulations. In the stable position, the energies obtained using the two {\bf k}-point meshes only differ by 3 meV.

The PIMD simulations are performed at T=100, 200, 300, 450, 600 and 800 K, in the NVT ensemble using a stochastic Langevin thermostat, and a time step of 5 atomic time units ($\approx$ 0.12 fs). The staging transformation is applied, and the number of imaginary time slices for the different temperatures studied is 64, 32, 22, 15, 11 and 8 (to maintain $P \times T =$ cte).

\begin{figure*}[htbp]
    {\par\centering
    {\scalebox{0.5}{\includegraphics{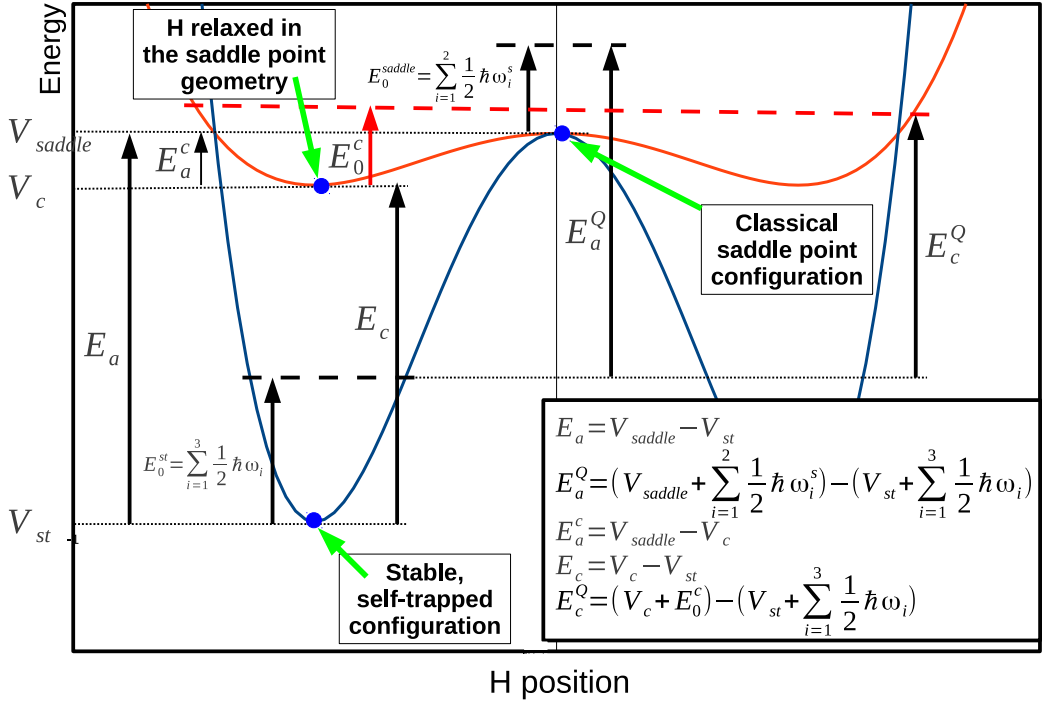}}}
    \par}
     \caption{{\small Schematic representation of the different configurations studied, with definitions of the energies involved in the problem. Only the adiabatic case is represented. Solid blue line: energy as a function of reaction coordinate with full relaxation of all the degrees of freedom. Solid red line: energy as a function of reaction coordinate with atoms frozen in the saddle point geometry (except H).
$V_{st}$: energy of the stable, self-trapped configuration;
$V_{saddle}$: energy of the classical saddle point configuration;
$V_{c}$: energy of the configuration with the position of H optimized in the fixed saddle point geometry (adiabatic coincidence configuration);
$E_a$: classical energy barrier;
$E_a^Q$: semi-classical energy barrier;
$E_0^{st}$: zero-point energy in the stable, self-trapped configuration;
$E_0^c$: zero-point energy in the coincidence configuration;
$E_a^c$: adiabatic coincidence barrier;
$E_c$: adiabatic coincidence energy, without accounting for zero-point energies;
$E_c^Q$: adiabatic coincidence energy, accounting for zero-point energies in the stable and coincidence configurations.
}}
    \label{definitions}
\end{figure*}

\section{Static configurations: stable, saddle point and coincidence}
\label{results}

The stable, saddle point and coincidence configurations (in the adiabatic case), with their energies, are schematically presented on Fig.~\ref{definitions}.

\subsection{Self-trapping, classical energy barrier for transfer and reorientation}
First, we optimize the configuration with the proton in its stable site (self-trapped), in the saddle point for transfer and in the saddle point for reorientation, and compute the classical barrier for transfer and for reorientation (Tab.~\ref{barriers}). The calculations are performed using the two Sn pseudopotentials (4-electron and 14-electron). We find as classical barrier for transfer (resp. reorientation) 0.32 eV (resp. 0.21 eV), using the 14-electron Sn atomic data. The 4-electron Sn atomic data provides very close values of 0.29 (resp. 0.22 eV). {\it In the classical regime, the transfer is thus the rate-limiting mechanism of protonic transport in BSO.}

\begin{table}[h]
\small
  \caption{\ Classical ($E_a$) and coincidence ($E_a^c$) energy barriers for protonic transfer and reorientation in barium stannate, using (i) a 14-electron pseudopotential for Sn, and (ii) a 4-electron pseudopotential for Sn. Both the adiabatic and non-adiabatic cases are considered.}
  \label{barriers}
  \begin{tabular*}{0.5\textwidth}{@{\extracolsep{\fill}}lll}
 \hline
         & Sn     &   Sn    \\
         & 14 electrons & 4 electrons  \\
 \hline
Transfer &  &  \\
Classical barrier $E_a$ &  0.32 & 0.29 \\
Coincidence barrier (adiab) $E_a^c$     &  0.05 & 0.04 \\
Coincidence barrier (non adiab) $E_a^c$ &   --  & 0.29 \\
 \hline
Reorientation &  &  \\
Classical barrier $E_a$ &  0.21 & 0.22 \\
Coincidence barrier (adiab) $E_a^c$ & 0.07 & 0.08 \\
Coincidence barrier (non adiab) $E_a^c$ & --  & 0.19 \\
 \hline
  \end{tabular*}
\end{table}

\begin{figure}[htbp]
    {\par\centering
    {\scalebox{0.40}{\includegraphics{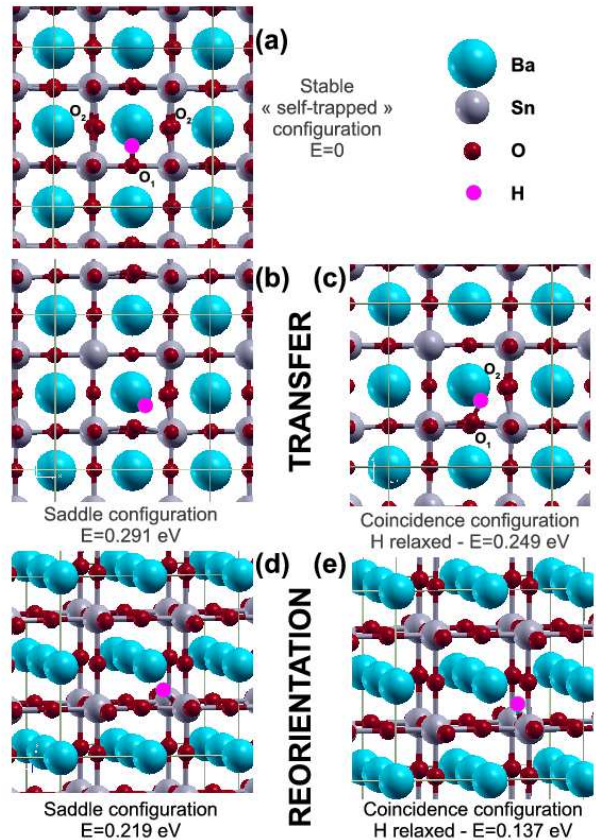}}}
    \par}
     \caption{{\small Atomic configurations: 
(a) Self-trapped configuration;
(b) Saddle point configuration for transfer;
(c) H$^+$ optimized in the saddle point geometry for transfer;
(d) Saddle point configuration for reorientation;
(e) H$^+$ optimized in the saddle point geometry for reorientation.}}
    \label{fig2}
\end{figure}

Self-trapping in the stable configuration is caused by lattice distortions around the proton, mainly those involving the two oxygens 2nd neighbors of H$^+$ (O$_2$): these two atoms are attracted by H$^+$, so that the Sn-O$_2$-Sn bond bends, forming an angle of 165$^{\circ}$ (Fig.~\ref{fig2} (a)). The two Sn cations 1st neighbors of H$^+$ are also slightly pushed away. The local polarization field resulting from such distortions is at the root of H self-trapping.

\subsection{Semi-classical regime: harmonic quantum corrections}

The semi-classical barrier, $E_a^Q$, is obtained by adding a harmonic quantum correction to the classical activation energy $E_a$. We make the approximation that the vibration modes of the host, i.e. that do not involve motions of H, are not significantly impacted by the position of H (stable or saddle point)~\cite{sundell2007}. Thus, only the (stable) modes involving H (3 in the stable position, 2 in the saddle position) provide significant contributions to the quantum correction:

\begin{equation}
E_a^Q = \{ V_{saddle} + \sum_{i=1}^{2} \frac{1}{2} h \nu_i^s \} - \{  V_{st} + \sum_{i=1}^{3} \frac{1}{2} h \nu_i \},
\end{equation}

$\{ \nu_i \}$ (resp. $\{ \nu_i^s \}$) is the set of the 3 (resp. 2 real) eigenfrequencies of the modes involving the motion of H in the stable, self-trapped (resp. saddle point) configuration. 
These eigenfrequencies are computed using a finite-difference method, by displacing the H atom by $\pm$~0.02~\AA{} in the three directions, starting from the stable or saddle point optimized geometries. Their values, and the associated zero-point energy, $\sum_i \frac{1}{2} \hbar \omega_i$ are given in Tab.~\ref{zero-point}.

As described above, the semi-classical regime occurs above $T_c$ (crossover temperature towards the quantum regime) and below the Debye temperature $T_D$. Using Gillan's approximate formula and the imaginary pulsations computed at the classical saddle positions (Tab.~\ref{zero-point}), we estimate $T_c$ $\sim$ 245 K (transfer) and $\sim$ 150 K (reorientation). Note that the operating temperature of PCFCs corresponds therefore to the semi-classical regime.

\begin{table}[h]
\small
  \caption{\ Eigenfrequencies (cm$^{-1}$) and corresponding energy quanta $\hbar \omega_i$ (meV) in the stable, coincidence and saddle point configurations for transfer and reorientation, and associated zero-point energies $\sum_{i} \frac{1}{2} \hbar \omega_i$ (meV), where $i$ runs over the stable vibration modes that involve the H atom (3 modes in stable sites, 2 modes at saddle point positions). In parenthesis are given the zero-point energies estimated by the path-integral simulation at low temperature.}
  \label{zero-point}
  \begin{tabular*}{0.5\textwidth}{@{\extracolsep{\fill}}llll}
 \hline
Configuration & Eigenfrequencies& $\hbar \omega$ & Zero-point energy \\
              &  (cm$^{-1}$)   & (meV)   &   (meV)   \\
\hline
Stable  & 3335&  413.5    &  313.6 \\     
(self-trapped)   & 935 &  116.0    &   \\
        & 788  & 97.7    &   \\
\hline
Transfer & & &  \\

Saddle point  & 1353   & 167.8    &  192.9    \\     
        & 1759   & 218.0    &           \\
       & 1073$i$ & 133.1$i$ &           \\

Coincidence  & 2190 & 271.5    &  270.3 ($\sim$ 240) \\     
(adiabatic)  & 1234 & 153.0    &   \\
             & 937  & 116.2    &   \\

Coincidence      & 2854 &  353.9    &  309.3  \\     
(non-adiabatic)  & 1101 &  136.5    &   \\
                 & 1035 &  128.3    &   \\

Coincidence      & 1761 &  218.3    &  193.2  \\     
(non-adiabatic)  & 1357 &  168.2    &   \\
(saddle point)   & 1566$i$ &  194.2$i$    &   \\

\hline
Reorientation & &  &  \\
Saddle point  & 3661  & 453.9 &   282.6   \\     
        & 899   & 111.4 &      \\
        & 650$i$ & 80.6$i$ &     \\

Coincidence  & 3406  & 422.3  & 311.5 ($\sim$ 290) \\     
(adiabatic)  &  888  & 110.1  &   \\
             &  730  &  90.5  &   \\

Coincidence      & 3418  &  423.8    & 314.4  \\     
(non-adiabatic)  & 843   &  104.5    &   \\
                 & 812   &  100.6    &   \\

Coincidence      & 3819  &  473.5    & 284.4  \\     
(non-adiabatic)  & 769   &  95.4    &   \\
(saddle point)   & 749$i$   &  92.8$i$    &   \\
 \hline 
  \end{tabular*}
\end{table}

In the stable, self-trapped, position, one recovers the three modes caracteristic of the vibration of the hydroxyl group: stretching (3335 cm$^{-1}$) and bending (935 and 788 cm$^{-1}$). The zero-point energy is 314 meV, roughly as high as the classical value of the transfer barrier. However, the relevant zero-point energy to be compared to the transfer barrier is the one associated with motion along the reaction coordinate. For transfer, this is the stretching mode, whose zero-point energy $\approx$ 207 meV is $\approx$ 70~\% of the barrier, enlighting the possible importance of anharmonic effects in the vibrationnal ground state of H in its stable site.

The semi-classical barriers $E_a^Q$ are given in Tab.~\ref{barriers-corr} and compared to the classical ones. As expected, they are lower (0.17 eV for transfer, 0.19 eV for reorientation), in line with the calculations of Sundell et al. on BZO~\cite{sundell2007}. {\it The quantum fluctuations, accounted for at the harmonic level, have a noticeable effect and significantly lower the barrier because the zero-point energies are larger in the stable site than in the saddle points}. The energy barrier reduction is stronger in the case of transfer ($\Delta E_{ZPE}$ = -0.12 eV) because the mode that softens is the stretching one, while it is a bending mode in the case or reorientation ($\Delta E_{ZPE}$ = -0.03 eV).

\begin{table}[h]
\small
  \caption{\ Classical barrier $E_a$, semi-classical barrier $E_a^Q$ and coincidence energy ($E_c$: without quantum correction, $E_c^Q$: including quantum correction) (eV) for proton transfer and reorientation in BSO.}
  \label{barriers-corr}
  \begin{tabular*}{0.5\textwidth}{@{\extracolsep{\fill}}llll}
 \hline
   &  Transfer  &  Reorientation  \\
\hline
Classical $E_a$          &  0.29  &  0.22    \\
Semi-classical $E_a^Q$   &  0.17  &  0.19    \\
\hline
Adiabatic case & & \\
Coincidence  $E_c$       &  0.25  &  0.14    \\
Coincidence $E_c^Q$      &  0.21  &  0.14    \\
(harmonic correction) & & \\
Coincidence $E_c^Q$      &  0.18  &  0.11    \\
(PIMD correction) & & \\
\hline
Non adiabatic case & & \\
Coincidence  $E_c$       &  0.18  &  0.09    \\
Coincidence $E_c^Q$      &  0.17  &  0.10    \\
 \hline
  \end{tabular*}
\end{table}

\subsection{Coincidence configuration and coincidence barriers: adiabatic case}

As described by Fukai~\cite{fukai} and Sundell et al.~\cite{sundell2007}, the coincidence configuration in the adiabatic picture is taken as that of the (classical) saddle point geometry~\cite{cherry1995,cherry1995b}. All the atoms of the supercell are thus kept fixed in the positions of this (classical) saddle point configuration, excepting H which is left free to relax in the resulting, frozen, coincidence potential $V^c_{ad}(\vec r)$. $V^c_{ad}(\vec r)$ roughly looks like a 3D double-well energy surface. The coincidence configuration in the adiabatic case (energy $V_{c}$) is thus obtained by optimizing the position of H$^+$in $V^c_{ad}(\vec r)$. The coincidence barrier, i.e. the barrier separating the two stable sites in the coincidence potential, is $E_a^c = V_{saddle} - V_{c}$. Of course, it is such that $E_a^c \leq E_a = V_{saddle} - V_{st}$, see Fig.~\ref{definitions}. The values of the coincidence barriers in the adiabatic case, as compared to the classical ones, are given in Tab.~\ref{barriers}.
We also compute the vibration modes of H in these coincidence configurations, for both mechanisms (Tab.~\ref{zero-point}).

\subsubsection{Transfer: }
The stable site in the coincidence configuration for transfer in the adiabatic case is shown on Fig.~\ref{fig2} (c). In this configuration, self-trapping originating from the atomic distortions is considerably weakened. In particular, the O$_1$-O$_2$ distance is reduced to 2.41~{\AA} (instead of 2.76~{\AA} in the self-trapped configuration), and the OH bondlength is stretched to 1.06~{\AA}. The distance of H to the final oxygen O$_2$ (hydrogen bond) is reduced to 1.46~{\AA} (instead of 2.12~{\AA} in the self-trapped configuration), reflecting a strong hydrogen bond H ... O$_2$. Consequently, the coincidence barrier is strongly lowered with respect to the classical one: 0.05 eV (0.04 eV using the 4-electron pseudopotential). $V^c_{ad}(\vec r)$ is very flat in the case of transfer.

Moreover, due to the strong hydrogen bond between H and O$_2$, the stretching mode is strongly softened to 2190 cm$^{-1}$ (to be  related to the very flat $V^c_{ad}(\vec r)$), and the zero-point vibrationnal energy of H is 270 meV (versus 314 meV in the stable self-trapped configuration). The adiabatic coincidence energy $E_c = V_c - V_{st}$ (Tab.~\ref{barriers-corr}) is 0.25 eV. Correcting this value from the (harmonic) zero-point energies in the stable self-trapped and coincidence configuration, we get $E_c^Q =$ 0.21 eV.

\subsubsection{Reorientation: }
The stable site in the coincidence configuration for reorientation is shown on Fig.~\ref{fig2} (e).
It corresponds to an OH bond not strictly parallel to an axis of the cubic structure, but partially turned, by $\approx$ 14.3$^{\circ}$, as if reorientation had started. The coincidence barrier for reorientation is 0.07 eV (0.08 eV using the 4-electron pseudopotential). This value is noticeably higher than the coincidence barrier for transfer.

Since the stretching mode keeps a high frequency at 3406 cm$^{-1}$, the zero-point vibrationnal energy of H remains as high as in the stable, self-trapped site (312 meV). The adiabatic coincidence energy $E_c = V_c - V_{st}$ (Tab.~\ref{barriers-corr}) is 0.14 eV. Correcting this value from the (harmonic) zero-point energies in the stable self-trapped and coincidence configuration, we get $E_c^Q =$ 0.14 eV.

For the two mechanisms, the coincidence barriers are thus much lower than the classical ones, in line with the similar calculation of Sundell et al.~\cite{sundell2007} performed on BZO. Moreover, their order is reversed: classically, the transfer has the highest energy barrier (0.29 eV versus 0.22 eV), while in their respective adiabatic coincidence configurations, this is the reorientation that exhibits the highest barrier (0.08 eV versus 0.04 eV).

\subsection{Coincidence configuration and coincidence barriers: non-adiabatic case}
In the non-adiabatic picture of thermally-assisted hydrogen tunneling, we take the coincidence configuration as that having the lowest energy under the following constraints:

(i) H remains on one single side of the barrier;

(ii) the environment of H, i.e. the set of all the atoms of the supercell except H, keeps the symmetry group of the saddle point configuration, making therefore the initial and final sites symmetry-equivalent.

This is equivalent to search for the most stable position of H in a perfectly symmetric environment. 
Practically, this configuration is reached by performing geometry optimization with forces on all the atoms except H symmetrized according to the symmetry operations of the space group of the saddle point configuration ($Amm2$ in the present case). These symmetries are determined and stored in a previous run.

As expected, this leads to a configuration more stable than the adiabatic one, but, of course, still less stable than the self-trapped one. The coincidence barriers are then determined by fixing these new atomic positions of the environment and relaxing the position of H in the symmetry plane separating the initial and final sites. The saddle point in this configuration is thus not the same as the classical saddle point. 
This leads to values of $E_a^c$ much higher than in the adiabatic case, rather close to the classical ones (Tab.~\ref{barriers}). We also compute the vibration modes of H in these non-adiabatic coincidence configurations and in the corresponding saddle points, for both mechanisms (Tab.~\ref{zero-point}).

\subsubsection{Transfer: }

In the coincidence configuration for transfer, the OH bondlength is 1.02 \AA, the hydrogen bond between H and O$_2$ is 1.73 \AA and the O$_1$-O$_2$ distance is 2.59 \AA. These three values are, as expected, exactly between the ones of the stable self-trapped and adiabatic coincidence configurations (Tab.~\ref{geometries}). Also, the stretching mode has its frequency at 2854 cm$^{-1}$, between the values of the stable (3335 cm$^{-1}$) and adiabatic coincidence (2190 cm$^{-1}$) configurations.

The coincidence barrier $E_a^c$ keeps a high value of 0.29 eV, equal to the classical barrier $E_a$, while the coincidence energy $E_c = V_c - V_{st}$ (in which $V_c$ is, here, the energy of the non-adiabatic coincidence configuration) is 0.18 eV (Tab.~\ref{barriers-corr}), naturally lower than in the adiabatic case. Correcting this value from the (harmonic) zero-point energies in the stable and non-adiabatic coincidence configurations, we get $E_c^Q =$ 0.17 eV. The zero-point correction does almost not modify $E_c$ because it has almost the same values in both cases.

\subsubsection{Reorientation: }

In the coincidence configuration for reorientation, the OH bondlength keeps the value characteristic of the hydroxyl group (0.985 \AA). It corresponds to an OH bond partially turned, once again, by $\approx$ 6.9$^{\circ}$, as if reorientation had started. This angle of rotation is nevertheless not as strong as in the adiabatic case (Tab.~\ref{geometries}).

The coincidence barrier $E_a^c$ keeps a high value of 0.19 eV, very close to the classical barrier $E_a =$ 0.22 eV, while the coincidence energy $E_c = V_c - V_{st}$ (in which $V_c$ is, here, the energy of the non-adiabatic coincidence configuration) is 0.09 eV (Tab.~\ref{barriers-corr}), naturally lower than in the adiabatic case. Correcting this value from the (harmonic) zero-point energies in the stable and non-adiabatic coincidence configurations, we get $E_c^Q =$ 0.10 eV. The zero-point correction does almost not modify $E_c$ because it has almost the same values in both cases.

Several characteristics (bondlengths, mode frequencies, energy) of the self-trapped, non-adiabatic and adiabatic coincidence configurations are gathered in Tab.~\ref{geometries}. They illustrate how self-trapping is progressively weakened from the self-trapped to the adiabatic configuration, the non-adiabatic one being inbetween.

\subsubsection{Transition rates: }
The coincidence energies in the non-adiabatic case are thus 0.17 ev (transfer) and 0.10 eV (reorientation). In the adiabatic case, the corresponding values are finally quite close: 0.18-0.21 eV (transfer) and 0.11-0.14 eV (reorientation). In the adiabatic case, the prefactor of the transition rate is $\sim$ 10$^{-13}$ s. In the non-adiabatic case, it depends on $J_0$ the bare tunneling matrix element of the proton. We give a rough estimate of $J_0$ using the approximate formula provided by Drechsler et al.~\cite{fukai,drechsler}, 
$J_0 = \frac{1}{2} \hbar \omega_0 e^{-\frac{m \omega_0 d^2}{4 \hbar}}$.

In the coincidence geometry for transfer, we have $\hbar \omega_0$ = 353.9 meV and $d$ = 0.76~\AA, which leads to $J_0$ = 8$\times$10$^{-4}$ meV.
Using the Flynn-Stoneham formula, we get a very weak transition rate prefactor of $\sim$ 2$\times$10$^{4}$ s$^{-1}$ at T=100 K and $\sim$ 1$\times$10$^{4}$ s$^{-1}$ at T=300 K. This is several orders of magnitude smaller than the prefactor expected in the case of adiabatic processes ($k_D \sim$ 10$^{13}$ s$^{-1}$). The small difference of coincidence energies between both cases does not compensate the difference in the prefactors: at T=100 K, we find $k$=6$\times$10$^{-5}$ s$^{-1}$ in the non-adiabatic case (using $E_c$=0.17 eV) versus $k$=2.6$\times$10$^{2}$ s$^{-1}$ in the adiabatic case using a coincidence energy of 0.21 eV.

In the reorientation case, the lower coincidence barrier in the non-adiabatic case (0.19 eV versus 0.29 for transfer) and the higher coincidence barrier in the adiabatic case (0.08 eV versus 0.04 for transfer) significantly reduce the difference (by $\sim$two orders of magnitude). We find, at T=100 K, $k$=2$\times$10$^1$ s$^{-1}$ (non-adiabatic case) and $k$=9$\times$10$^5$ s$^{-1}$ (adiabatic case). Here the tunneling matrix element $J_0$ remains weak ($\sim$ 7$\times$10$^{-3}$ meV) because the energy barrier is wide ($d$=1.22 \AA).

These estimations provide arguments in favor of the adiabatic picture.

\begin{table}[h]
\small
  \caption{\ Evolution of geometric parameters and mode frequencies from the stable configuration to the non-adiabatic coincidence and adiabatic coincidence configurations. O$_1$H is the hydroxyl bondlength and H...O$_2$ is the hydrogen bond between the proton and O$_2$. The energies of the various configurations are given with respect to the stable, self-trapped one taken as reference.}
  \label{geometries}
  \begin{tabular*}{0.5\textwidth}{@{\extracolsep{\fill}}llll}
 \hline
   &  Stable        &  Coincidence  & Coincidence  \\
   &  self-trapped  &   non-adiabatic    &  adiabatic   \\
\hline
Transfer & & & \\
O$_1$H  (\AA)    & 0.99   & 1.02  & 1.06  \\
H...O$_2$ (\AA)  & 2.12   & 1.73  & 1.46  \\
O$_1$-O$_2$(\AA) & 2.76   & 2.59  & 2.41  \\
Stretching mode (cm$^{-1}$)&  3335  & 2854 & 2190 \\
Energy (eV)    &  0  &  0.18  &  0.25  \\
\hline
Reorientation & & & \\
Angle of rotation ($^{\circ}$) & 0 & 6.9 & 14.3  \\
Energy (eV)   &   0    & 0.09  &  0.14  \\
 \hline
  \end{tabular*}
\end{table}


\section{Path-Integral description of H in the adiabatic coincidence configuration}
\label{pi}

Therefore, we now focus on the adiabatic case. If the adiabatic picture is valid, as suggested by the previous estimations, 
it means that the proton has the time to equilibrate its motion in the fixed potential of the adiabatic coincidence configuration. Thus, we use the PIMD technique to describe the state of the proton in the (adiabatic) coincidence configuration for transfer and for reorientation, i.e. the classical saddle point geometry. The atoms surrounding H are kept fixed in this configuration, and the PIMD equations of motion are applied to the sole proton. Trajectories of several hundreds to several thousands steps are performed.

\subsection{Density of probability of a thermalized proton in the coincidence geometry}

From the PIMD trajectories, we extract the densities of probability $P(\lambda)$ of the proton as a function of a parameter $\lambda$ describing its position along the diffusion path.

\subsubsection{Transfer: }

\begin{figure}[htbp]
    {\par\centering
    {\scalebox{0.3}{\includegraphics{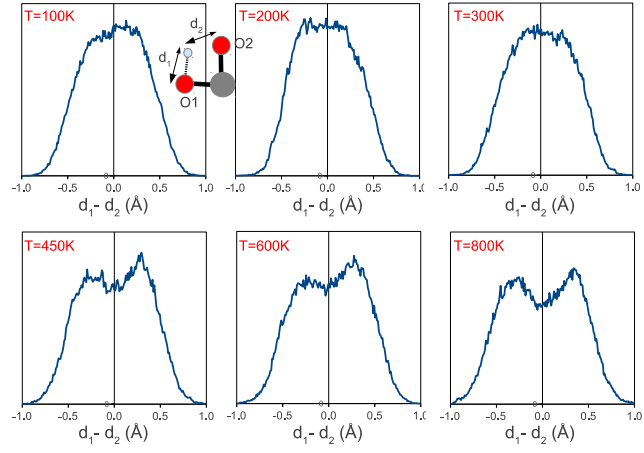}}}
    \par}
     \caption{{\small Density of probability of $\lambda = O_1H - O_2H$, for the proton thermalized in the frozen adiabatic coincidence geometry for transfer (classical saddle point).}}
    \label{fig3}
\end{figure}

In the case of transfer, we choose $\lambda = O_1H - O_2H$ (Fig.~\ref{fig3}). The H atom evolves in a very flat energy landscape ($E_a^c$ = 0.04 eV). Two different behaviors are observed:

(i) for $T \leq$ 300 K, the density of probability has a single, broad, maximum centered at $\lambda$ = 0. This is characteristic of a quantum behavior because a classical particle would have its maximum of probability in the potential minima. The ground state of H in the adiabatic coincidence geometry for transfer has thus a single maximum at the saddle point ($\lambda$ = 0). We will explain hereafter that it corresponds to a zero-point energy higher than the coincidence barrier;

(ii) for $T \geq$ 450 K, the density of probability has two maxima, for $\lambda = \pm \lambda_0$. This is characteristic of a semi-classical behavior. The positions of these two maxima tend to those of the potential minima as T increases, as the classical limit is approached.

For H evolving in the frozen adiabatic coincidence potential, the transition from a quantum to a semi-classical behavior occurs between 300 K and 450 K. 
The single-peaked distribution below 300 K suggests that H is frozen in its ground state and that the zero-point energy exceeds the coincidence barrier, corresponding to a {\it quantum} over-barrier motion.
Above such temperature, the excited states of H start to have significant contributions,
leading to a {\it semi-classical} over-barrier motion.

\subsubsection{Reorientation: }

\begin{figure}[htbp]
    {\par\centering
    {\scalebox{0.3}{\includegraphics{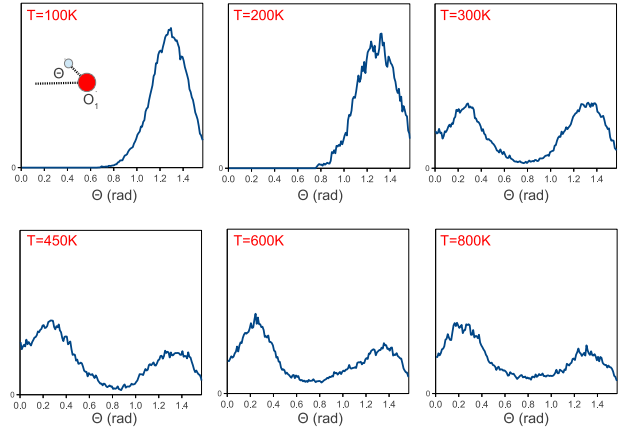}}}
    \par}
     \caption{{\small Density of probability of reorientation angle $\theta$ (in radians), for the proton thermalized in the frozen adiabatic coincidence geometry for reorientation (classical saddle point). For T=100 and 200 K, the trajectories are non ergodic.}}
    \label{fig4}
\end{figure}

In the case of reorientation, the situation is radically different, because the coincidence barrier is significantly larger (0.08 eV). The equilibrium density of probability is plotted on Fig.~\ref{fig4} as a function of $\Theta$ (angle between the projection of OH onto the rotation plane, and its initial direction, in the stable site).

At low temperature, the proton is found trapped in one single well. Above 300 K, however, delocalization over the two wells is observed, with a double-peaked density of probability. Note that, since the potential is symmetric with respect to the saddle point position $\lambda$=0, $P(\lambda)$ should exhibit a symmetric profile at any temperature. The non-symmetric profile found at T=100 and 200 K reflects lack of ergodicity. Longer trajectories, if possible, would obviously provide the expected double-peaked profile.

\subsection{Quantum behavior of the proton in the adiabatic coincidence geometry}

\subsubsection{Spatial extension of the thermalized proton: }

\begin{figure}[htbp]
    {\par\centering
    {\scalebox{0.3}{\includegraphics{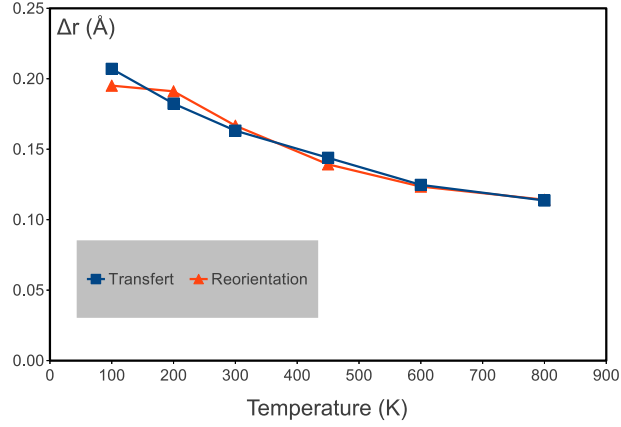}}}
    \par}
     \caption{{\small Quantum spatial spread of the protonic wave packet, $\Delta r$ (\AA), as a function of temperature (K), for a proton thermalized in the adiabatic coincidence geometry for transfer (blue symbols) and reorientation (orange symbols).}}
    \label{deltar}
\end{figure}

We compute the spatial extension of the protonic wave packet (Fig.~\ref{deltar}), $\Delta r$, as

\begin{equation}
\Delta r^2 = < \frac{1}{P} \sum_{s=1}^{P} || \vec r^{(s)}  -  \vec r^c ||^2  >,
\end{equation}

in which $\vec r^c = \frac{1}{P} \sum_{s=1}^P \vec r^{(s)}$ is the centroid associated to the proton. 

$\Delta r$ decreases with temperature, from $\approx$ 0.2~{\AA} at T=100 K to 0.11~{\AA} at T=800 K. The large fluctuation at low temperature is characteristic of the quantum nature of the protonic motion. The slow decrease of $\Delta r$ with temperature reflects a slow quantum/classical transition. At T=800 K, in the semi-classical regime, $\Delta r$ keeps a quite large value of 0.11~{\AA}.

\subsubsection{Energy: }

From the Path-Integral simulations, we compute the energy of the proton by using the Path-Integral Virial estimator:

\begin{equation}
\nonumber
E^{(vir)} = \frac{3}{2} k_B T + \frac{1}{P} < \sum_{s=1}^{P} \frac{1}{2} (\vec r^{(s)} - \vec r^c).(-\vec F^{(s)}) > + \frac{1}{P} < \sum_{s=1}^P V(\vec r^{(s)} )>,
\end{equation}

\begin{figure}[htbp]
    {\par\centering
    {\scalebox{0.325}{\includegraphics{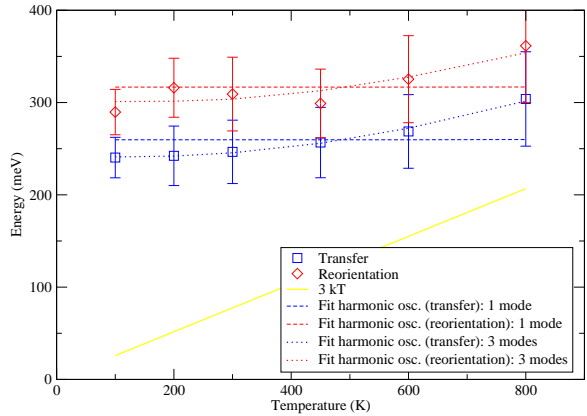}}}
    \par}
     \caption{{\small Energy (meV) versus temperature (K) of the proton thermalized in the coincidence potential for transfer (blue symbols) and reorientation (red symbols). }}
    \label{energy}
\end{figure}

Fig.~\ref{energy} plots this mean energy as a function of temperature, both for transfer and reorientation.

The zero-point energy of the proton in the coincidence potential can be extracted from an extrapolation at T=0 K: $\approx$ 240 meV for transfer, $\approx$ 290 meV for reorientation. A fit of the data by the energy of harmonic oscillators ($E = \sum_i \frac{1}{2} e_i + \frac{e_i}{e^{e_i / k_B T} - 1}$) leads to a zero-point energy $\sum_i \frac{1}{2} e_i$ = 260 (317) meV for transfer (reorientation) in the case of one mode, and 241 (301) meV for transfer (reorientation) in the case of three modes
(dashed curves on Fig.~\ref{energy}).
A classical particle thermalized in a 3D harmonic potential would have a mean energy of $3 k_B T$. This quantity is plotted for comparison in Fig.~\ref{energy}. At T=100~K, the energy of the proton largely exceeds $3 k_B T$, in both the transfer (by $\sim$ 215~meV) and reorientation (by $\sim$ 265~meV) cases.

These estimations of the zero-point energy in the adiabatic coincidence potential allow to give better estimation for the coincidence energy, including quantum zero-point energies in the stable and coincidence configurations, $E_c^Q = [V_{c}+ E_0^c] - [V_{st} + E_0^{st}]$, with $E_0^{st} = \sum_{i=1}^{3} \frac{1}{2} \hbar \omega_i$ = 314 meV, and $E_0^c$ the zero-point energy deduced from the Path-Integral simulations (240 meV for transfer, 290 meV for reorientation). We find $E_c^Q$= 0.18 eV for transfer, 0.11 eV for reorientation (Tab.~\ref{barriers-corr}).

\section{Discussion}
\label{discussion}

\subsection{Interpretation of the adiabatic mechanism using a simple, one-dimensional quartic model}
\label{quartic}

In this part, we model the adiabatic coincidence potential by a one-dimensional quartic double well $V(x)=V_0 (\frac{x}{a} -1)^2(\frac{x}{a} +1)^2$. To mimick as much as possible the real situation, the double well barrier $V_0$ is identified with the coincidence barrier $E_a^c$, while $a$ is set to the distance separating the H$^+$ positions in the saddle point and in the stable configuration (in the adiabatic coincidence geometry). The parameters are given in Tab.~\ref{values-model}. Then, we compute the eigenstates of a quantum particle of mass $m$ confined in $V(x)$, by numerically solving the 
Schr\"odinger equation, $-\frac{\hbar^2}{2m} \frac{d^2 \phi}{dx^2}(x) + V(x) \phi(x) = E \phi(x)$, using a finite-difference method and exploiting the fact that the potential is a pair function. In particular, the ground state wave function $\phi_0(x)$ is obtained for the value of $E$ that provides a wave function (i) with no node, and (ii) satisfying $\phi'_0(0)$=0. 

\begin{table}[h]
\small
  \caption{\ Values of $V_0$ (eV) and $a$ (\AA) used to define the double well potential.}
  \label{values-model}
  \begin{tabular*}{0.5\textwidth}{@{\extracolsep{\fill}}llll}
 \hline
Mechanism   &  $V_0$ (eV)  &  $a$ (\AA)  \\
\hline
Transfer      &  0.042  &  0.2108    \\
Reorientation &  0.082  &  0.5187    \\
 \hline
  \end{tabular*}
\end{table}

\begin{figure}[htbp]
    {\par\centering
    {\scalebox{0.35}{\includegraphics{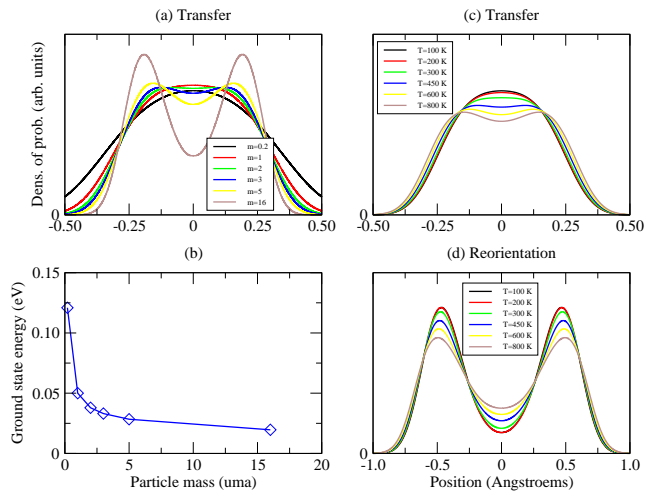}}}
    \par}
     \caption{{\small 
(a) Density of probability of the ground state for different particle masses in the case of transfer.
(b) Energy (eV) of the ground state as a function of particle mass in the case of transfer.
(c) Density of probability for the proton at finite temperature (transfer).
(d) Density of probability for the proton at finite temperature (reorientation). }}
    \label{model}
\end{figure}

\subsubsection{Ground state wave function as a function of particle mass in the case of transfer: }

First, we artificially vary the quantum character of the particle by varying its mass, from 0.2 to 16, in the case of transfer. The proton and the deuteron correspond to $m$=1 and $m$=2. The ground state energies increase with decreasing mass of the particle, as shown on Fig.~\ref{model} (b). The densities of probability in the ground state are shown in Fig.~\ref{model} (a). According to the particle mass, i.e. to the ground state energy E$_0$, different regimes are observed:

\begin{enumerate}

\item for $E_0 \leq V_0$ ($m$=16, 5, 3 and 2), the wave function exhibits two maxima, with a density of probability which is not zero at the saddle point (tunneling), and that increases with decreasing mass. For large mass, the maxima are well pronounced and narrow;

\item for $E_0 \geq V_0$ ($m$=1), the ground state progressively becomes insensitive to the details of the potential smaller than the ground state energy, and thus the barrier is not seen any more. The ground state wave function becomes therefore single-peaked, with its maximum localized at the saddle point $x$=0. Such regime has been sometimes called  "quantum localization"~\cite{kitamura,muons}. Beyond some high value of $E_0$ ($m$=0.2), the system is only sensitive to the confinement potential, and the wave function, single-peaked, spreads further and further towards $x>0$ and $x<0$. 

\end{enumerate} 

In the adiabatic potential for transfer, the proton ($m$=1) belongs to the second highly quantum regime, whereas the deuteron ($m$=2) belongs to the first one.

\subsubsection{Density of probability as a function of temperature: }

\begin{figure}[htbp]
    {\par\centering
    {\scalebox{0.36}{\includegraphics{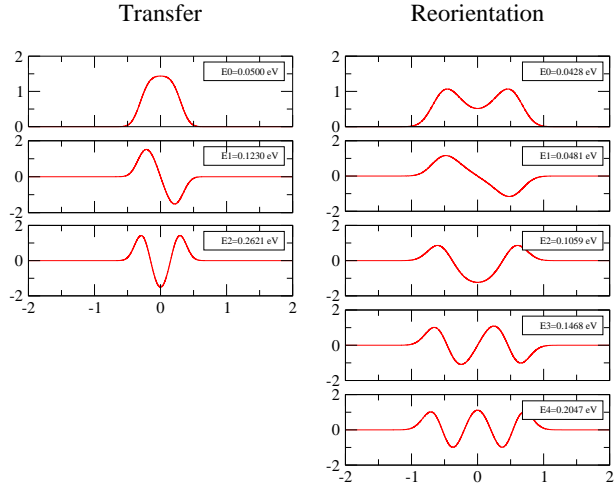}}}
    \par}
     \caption{{\small 
Wave functions of the first excited states of a proton in a 1D quartic double well, in the case of transfer ($E_0 > V_0$), and reorientation ($E_0 < V_0$).}}
    \label{wave_function}
\end{figure}

We now compute the probability distribution of the proton ($m$=1) in the double well, at finite temperature, from T = 100~K to T = 800~K. To this aim, we calculate the wave functions $\phi_n(x)$ and energies $E_n$ of the first excited states up to $E_n \sim$ 0.20 eV (see Fig.~\ref{wave_function}). The equilibrium density of probability, at finite $T$, is then calculated using the canonical probabilities, as

\begin{equation}
n(x;T) = \sum_{i=0}^{N_{st}} n_i(x) \frac{e^{-E_i / k_B T}}{Z}~~~ Z = \sum_{i=0}^{N_{st}} e^{-E_i / k_B T},
\end{equation}

the summation being truncated at $N_{st}$=3 (transfer) and 5 (reorientation) instead of $+\infty$.
$n_i(x) = |\phi_i(x)|^2$ is the density of probability in the $i^{th}$ excited state.

The densities obtained at different temperatures are shown in Fig.~\ref{model} (c) and (d). At low temperature, the particle is frozen in its ground state, its density of probability exhibits therefore a single maximum for $x$=0 in the case of transfer, and a double peak in the case of reorientation. The single peak in the case of transfer is due to the fact that the ground state energy exceeds the energy barrier: the quantum particle is quasi-insensitive to the details of the potential smaller than its ground state energy, and the shape of its wave function is mainly controlled by the confinement potential~\cite{note}.

Upon increasing temperature, the relative weight of the first excited state, which exhibits two extrema for non-zero $x$, increases, and the density of probability becomes progressively double-peaked in the case of transfer, precisely between 300 and 450 K. At very high temperature, in the classical limit, $n(x;T)$ would tend to the Boltzmann distribution, which is simply proportional to $e^{- V(x)/k_B T}$ and would have two maxima at $x=\pm a$. 

The present model, despite its simplicity, mimicks thus very well the behavior of H$^+$ in the true adiabatic coincidence potential, as found by the Path-Integral simulations (Sec.~\ref{pi}, Figs.~\ref{fig3} and ~\ref{fig4}).

\subsection{Adiabatic and non-adiabatic processes}
\label{adiabatic}

In the thermally-assisted quantum regime, transfer and reorientation have thus different characteristics. Here we tentatively estimate the typical time scale for tunneling in the adiabatic coincidence potential using quantum mechanics, for the two mechanisms. Comparison of these times with the lifetime of the coincidence state should allow to validate, or not, the adiabatic picture for each of the two mechanisms.

Determining unambiguously whether H$^+$ behaves adiabatically or not is a very challenging issue, because our PIMD simulations do not give access to any time-correlation function. More sophisticated methods, such as Centroid Molecular Dynamics should be employed, but require high computational cost.

Instead, we only provide here semi-quantitative arguments based on the previous quartic model. We estimate $\Delta t$, the characteristic time needed for the protonic wave packet to spread over the double well, starting from a state that is localized on one side of the barrier. In the coincidence configuration, such state is not an eigenstate of the Hamiltonian (all the eigenstates should have symmetric density of probability): it is a linear combination of the eigenstates, with typical energy extension $\Delta E$. The time needed for the wave packet to spread is provided by the time-energy uncertainty relation $\Delta t . \Delta E \sim \hbar$. Of course, $\Delta E$ depends on the kind of localization. Perfect localization at one point imply infinite $\Delta E$ and thus $\Delta t$=0. In order to make sense, we thus consider the minimal $\Delta E$ that allows to make a state localized on one side of the barrier.

For reorientation, localized states can easily be obtained as linear combinations of the ground state $| \phi_0 >$ and of the first excited state $|\phi_1 >$, as $( | \phi_0 > + |\phi_1 >)/\sqrt{2}$ and $( | \phi_0 > - |\phi_1 >)/\sqrt{2}$. Thus $\Delta E \sim E_1 - E_0 \sim$ 0.0053 eV and $\Delta t \approx$ 1.2$\times$10$^{-13}$ s.

For transfer, the peculiar form of the ground state, with a maximum at $x$=0, implies that states localized on one side of the
barrier have very large energy extension $\Delta E$. At least the first 3 states are needed ($\Delta E \sim E_2 - E_0 \sim$ 0.21 eV), leading to a typical time scale $\Delta t \approx$ 3.0$\times$10$^{-15}$ s for significant evolution of the protonic wave packet.

For the surrounding atoms, the shortest characteristic time is that associated with the highest phonon dispersion curve of barium stannate. Using phonon dispersion curves of BaSnO$_3$ previously computed~\cite{bevillon2007}, the typical period of the vibration modes of the highest-energy branch can be estimated at $\sim$ 5.0$\times$10$^{-14}$ s (it involves a collective motion of the oxygen atoms). The phonon modes associated with other branches have typical times $\geq$ 8$\times$10$^{-14}$ s, and the main peak is around 1.0$\times$10$^{-13}$ s. The "lifetime" of the coincidence configuration can therefore be estimated at $\sim$ 5.0$\times$10$^{-14}$ s - 1.0$\times$10$^{-13}$ s.

These estimations give support for the adiabatic approximation in the case of transfer, as inferred in Sec.~\ref{results}. 
In the case of reorientation, the typical time associated with the proton dynamics is roughly the same order of magnitude as the one for coincidence, suggesting that reorientation is a complex process, that can be completely described, neither by the fully non-adiabatic picture (Flynn-Stoneham), nor by the adiabatic one.

\subsection{Impact of exchange-correlation energy functional}

All our first-principles calculations have been performed in the framework of the GGA-PBE functional.
A major drawback of this functional in the case of protonic transfer is to strongly underestimate the energy barrier~\cite{sundell2007,hermet2013,barone1996}. The energy barriers $E_a$, $E_a^Q$, $E_c$ and $E_c^Q$ are thus probably underestimated in the present work. 

Higher barriers would correspond to a smaller tunneling probability, and thus higher $\Delta t$.
However, we point out that the zero-point energy computed with PBE in the coincidence configuration for transfer (restricted to the motion along the reaction coordinate) is $\sim$ 0.14 eV (harmonic approximation -- adiabatic case) or $\sim$ 0.11 eV (PIMD -- adiabatic case)
\footnote{
this energy is obtained by substracting to the PIMD energy the zero-point energies obtained in the harmonic approximation for the two bending modes, which are roughly orthogonal to the reaction coordinate
}. 
This value is much larger than the 0.04 eV coincidence barrier. Thus, even a 50\% underestimation of the barrier by PBE would still correspond to a ground state energy larger than the barrier. Thus, barrier underestimation by GGA-PBE has probably little influence on the nature of the transition state (with the maximum of probability at the saddle point) in the case of adiabatic transfer.

\subsection{Analogy with ice under pressure}

We have determined that, at temperatures typically lower than 300 K, 
when thermal fluctuations of the atoms in the surrounding matrix produce adiabatic coincidence configurations,
the proton occasionally evolves in a very flat energy landscape (typically during a few 10$^{-13}$ s). In this potential, we have seen that the protonic state -- that can be interpreted as the transition state for transfer -- exhibits strong quantum behavior.

Indeed, the barrier $E_a^c$ is so low (0.04 eV) that the ground state, in this adiabatic coincidence potential, is single-peaked and has its maximum of probability at the saddle point. Occurrence of such protonic states is dependent on the ability for the proton to behave adiabatically during the typical time for coincidence, but we have seen that arguments in favor of an adiabatic behavior do exist in the case of transfer.

This very peculiar state of the proton does occasionally exist in proton-conducting oxides at ambient pressure, but similar state has already been observed, at thermal equilibrium, in ice under high pressure~\cite{benoit1998,morrone2009}. In such system, the pressure can be used to control the distance between the neighboring oxygen atoms, sites between which the protons can jump. At low pressure, the protons behave classically, i.e. they remain localized at one oxygen site (ice VIII), forming a well-identified OH group hydrogen-bonded to another oxygen, in an asymmetric geometry (O-H ... O). Upon increasing pressure, a new phase (ice VII) appears with protonic disorder related to the possibility for the proton to tunnel through the barrier, leading to a symmetric probability distribution and the loss of well-identified OH groups in the structure. When the pressure is further increased, the energy barrier becomes finally so low that the protonic distribution looses its bimodal character and evolves to a single-peaked function with its maximum at the saddle point (ice X). The bimodal/unimodal character of the protonic distribution in that case is related to the oxygen-oxygen distance, and thus to the energy barrier for the proton to diffuse from an oxygen to the facing one. External pressure controls such distance. The transition state for protonic transfer in barium stannate is thus analogous to the state of H$^+$ in ice X.

In the case of reorientation, the barrier $E_a^c$ is larger (0.08 eV) so that the protonic ground state in the adiabatic coincidence potential has a bimodal density of probability. However, it is probable that the proton has not the time to behave adiabatically in that case, so that it remains localized on one side of the barrier during the time of the coincidence event. This prevents from making an analogy between the transition state for reorientation and the state of H$^+$ in ice VII.

We should add also here that quantum tunneling associated to loss of well-defined OH bond has also been suggested to occur in water monolayers adsorbed on transition metal surfaces~\cite{li2010}.

\section{Conclusion}

In this work, hydrogen diffusion in barium stannate has been investigated by first-principles density-functional calculations, in the classical, semi-classical and thermally-assisted quantum regimes.

From structural optimizations and frozen-phonon calculations, the energy barriers relevant in the classical (high-temperature) and semi-classical (intermediate temperature) regimes have been computed. In the classical regime, transfer appears as the rate-limiting phenomenon, while in the semi-classical one (above $\sim$ 150-250 K), the harmonic quantum corrections make the activation energies for transfer and for reorientation similar (quantum effects on the hopping barrier are stronger).
Correct characterization of this regime is particularly important since it corresponds to the operating temperatures of PCFCs.

In the phonon-assisted quantum regime (below $\sim$ 150-250 K), H diffusion is a complex process for which a simple reaction coordinate can not be simply identified, since it is related to modifications of the atomic environment of H, that occasionally weaken self-trapping by the means of thermal fluctuations. Both a non-adiabatic and an adiabatic behavior have been examined.
Simple estimations of the transition rate rather provide arguments in favor of an adiabatic behavior.

In the adiabatic coincidence configuration, the energy barriers for transfer and for reorientation are very low (0.04 and 0.08 eV). 
The state of the H atom evolving in this adiabatic coincidence potential has been simulated using the Path-Integral technique, 
assuming that H has the time to thermalize in this fixed potential (i.e. an adiabatic behavior).
It provides the density of probability and energy of the proton in the coincidence potential, under such assumption. These calculations reveal the highly quantum nature of H in this transition state, related to the very flat coincidence potential and the light mass of H.

In the case of transfer, the coincidence energy barrier is so low ($\approx$ 0.04-0.05 eV) that the ground state of H exhibits a single maximum, localized at the saddle point. This transition state of H is similar to the one observed in ice under very high pressures (ice X). In the case of reorientation, the coincidence energy barrier is higher (0.07-0.08 eV), and the ground state of H exhibits two maxima in the initial and final sites. However, our study suggests that an adiabatic description of the phonon-assisted quantum regime might be relevant only for transfer. For reorientation, the typical time scales associated with the existence of the coincidence configuration and the motion of H are roughly equal, making an adiabatic behavior less probable.

In both cases, the models that currently describe the thermally-assisted tunneling regime in metals, which are based on a fully non-adiabatic picture for the hydrogen motion (tunneling period much larger than typical time for coincidence) cannot be generalized to proton-conducting oxides. Extension of such studies to lower-symmetry compounds such as BaCeO$_3$ wouls be valuable~\cite{hermet2014}.

Besides, from our calculations, it is clear that protons in oxides behave semi-classically in the range of temperatures from $\sim$ 300 K to at least $\sim$ 1000 K, including the operation temperature of fuel cells. Then, any DFT calculations on proton conductors should include zero-point energies to give a precise description of proton diffusion mechanisms. In particular, we have shown here that such corrections are particularly relevant to identify the limiting mechanism among transfer and reorientation, that changes between the semi-classical and the classical regime.

\section{Acknowledgements}
This work was performed using the HPC resources of the TERA-100 supercomputer of CEA/DAM and from GENCI-CCRT/CINES (Grants 2013-096468 and 2014-096468). 
The authors acknowledge Professor T. Norby and T. Bj\o rheim for careful reading of the manuscript and providing valuable comments.
This work was supported by the Agence nationale de la Recherche through the IDEA-MAT project (ANR-12-BS08-0012-01).



\begin{thebibliography}{References}
\bibitem{note}
Let us consider a particle merged in a 1D potential $V(x)$, with eigenstates $\phi_i(x)$ and eigenenergies $E_i$. Rewriting the Schr\"odinger equation as 
$-\frac{\hbar^2}{2m}  \phi''(x) = (E_i - V(x)) \phi_i(x)$ shows that inflection points of the eigen wave function correspond, either to the nodes ($\phi_i(x)=0$), or to the points at which the potentiel is equal to the eigenenergy ($V(x)=E_i$). Let us focus on the ground state $\phi_0(x)$. Since $\phi_0(x)$ has no node, its inflection points are the $x$ for which $V(x)=E_0$.

For a partice merged in a double well potential, two cases must be considered:

(i) if $E_0$ is lower than the energy barrier separating the wells, there are 4 inflection points, leading to a ground state with two maxima of the density of probability, on both sides of the barrier;

(ii) if $E_0$ is higher than the energy barrier, there are only 2 inflection points. Since $\phi_0(x)$ has no node, it can be arbitrarily chosen as $> 0$. Inside the potential well, the second derivative $\phi_0''(x) < 0$ and thus $\phi_0'(x)$ is strictly decreasing and can thus vanish only once. Thus $\phi_0(x)$ will have a single maximum. If $V(x)$ is a pair function, this maximum is at zero, by symmetry.
\end{thebibliography}
\end{document}